\input{aipcheck}

\documentclass[
    ,final            
  ]
  {aipproc}

\layoutstyle{6x9}

\newcommand{\peg}{P\'EGASE}
\newcommand{\clo}{CLOUDY}

\newcommand{\mic}{\mbox{$\mu$m}}
\newcommand{\ngc}[1]{NGC$\,$#1}

\newcommand{\iizw}{II$\,$Zw$\,$40}
\newcommand{\hen}{He$\,$2-10}

\newcommand{\hii}{H$\,${\sc ii}}

\newcommand{\neiii}{Ne$\,${\sc iii}}
\newcommand{\neii}{Ne$\,${\sc ii}}
\newcommand{\siv}{S$\,${\sc iv}}

\newcommand{\neiiiline}{[\neiii] $\lambda 15.56\,\mic$}
\newcommand{\neiiline}{[\neii] $\lambda 12.81\,\mic$}
\newcommand{\sivline}{[\siv] $\lambda 10.51\,\mic$}

\newcounter{textlistctr}
\newcommand{\thetextlist}{, }
\newcommand{\textlist}[1]
           {\setcounter{textlistctr}{1}
            \renewcommand{\thetextlist}
            {{\it (\roman{textlistctr})}\stepcounter{textlistctr}}#1}

\graphicspath{{Figures/}}

\begin{document}

\title{Constraining the local radiation field and the grain size distribution 
       in dust SED modelling of dwarf galaxies}

\author{Fr\'ed\'eric {\sc Galliano}}{
  address={SAp, CEA/Saclay, l'Orme des Merisiers, 91191 Gif, France \\
           Infrared Astrophysics branch, code 685, NASA GSFC, Greenbelt MD 
              20771, USA}
  }

\begin{abstract}
I present a simple self-consistent dust spectral energy distribution (SED) 
model that has been applied to fit the well-sampled observed UV-to-radio SED of
four nearby starbursting dwarf galaxies.
The main originality of this model is that numerous multi-wavelength 
observations, from UV to millimeter (mm), constrain in a self-consistent 
manner, both the local radiation field and the grain size distribution.
I finally present the results of our model and discuss the average dust 
properties in these dwarf galaxies.
\end{abstract}

\maketitle


\section{Introduction: the difficulties to interpret an observed dust SED}

Most of the current dust models 
\citep{desert+90,li+01,draine+01,zubko+04} have been
developed to describe the emission from the diffuse Galactic ISM.
Consequently, their application to other galaxies is not straightforward.

First, the dust abundances and grain size distribution are
likely to depend significantly on the local physical conditions.
Indeed, the grains can undergo coagulation and accretion of material in dense 
media; fragmentation and erosion in diffuse shocked media; evaporation
of smaller grains in \hii\ regions and injection of newly-formed grains by 
evolved stars and supernovae.
Second, the interpretation of an observed IR-to-millimeter SED of a galaxy in 
terms of dust properties (composition, mass, size distribution) is 
difficult due to the fact that the hardness and intensity of the local 
interstellar radiation field (ISRF) vary strongly from quiescent to 
starforming regions.
This ISRF being the heating source of the dust, its spectral shape 
and intensity have a direct effect on the spectrum emitted by the dust.

\begin{figure}[htbp]
  \centering
  \begin{tabular}{cc}
    \includegraphics[width=0.5\linewidth]{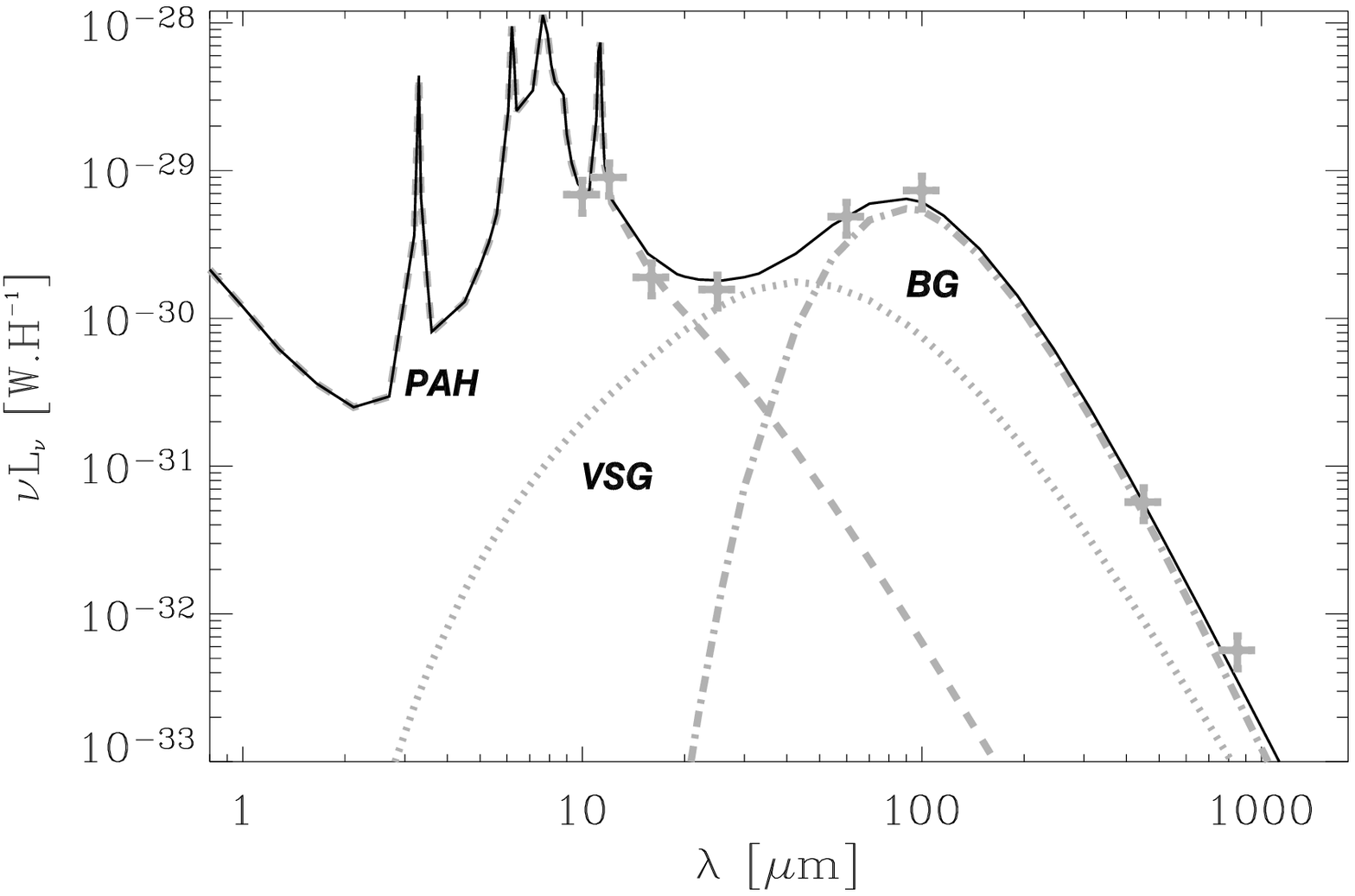}  & \hspace*{-0.6cm}
    \includegraphics[width=0.5\linewidth]{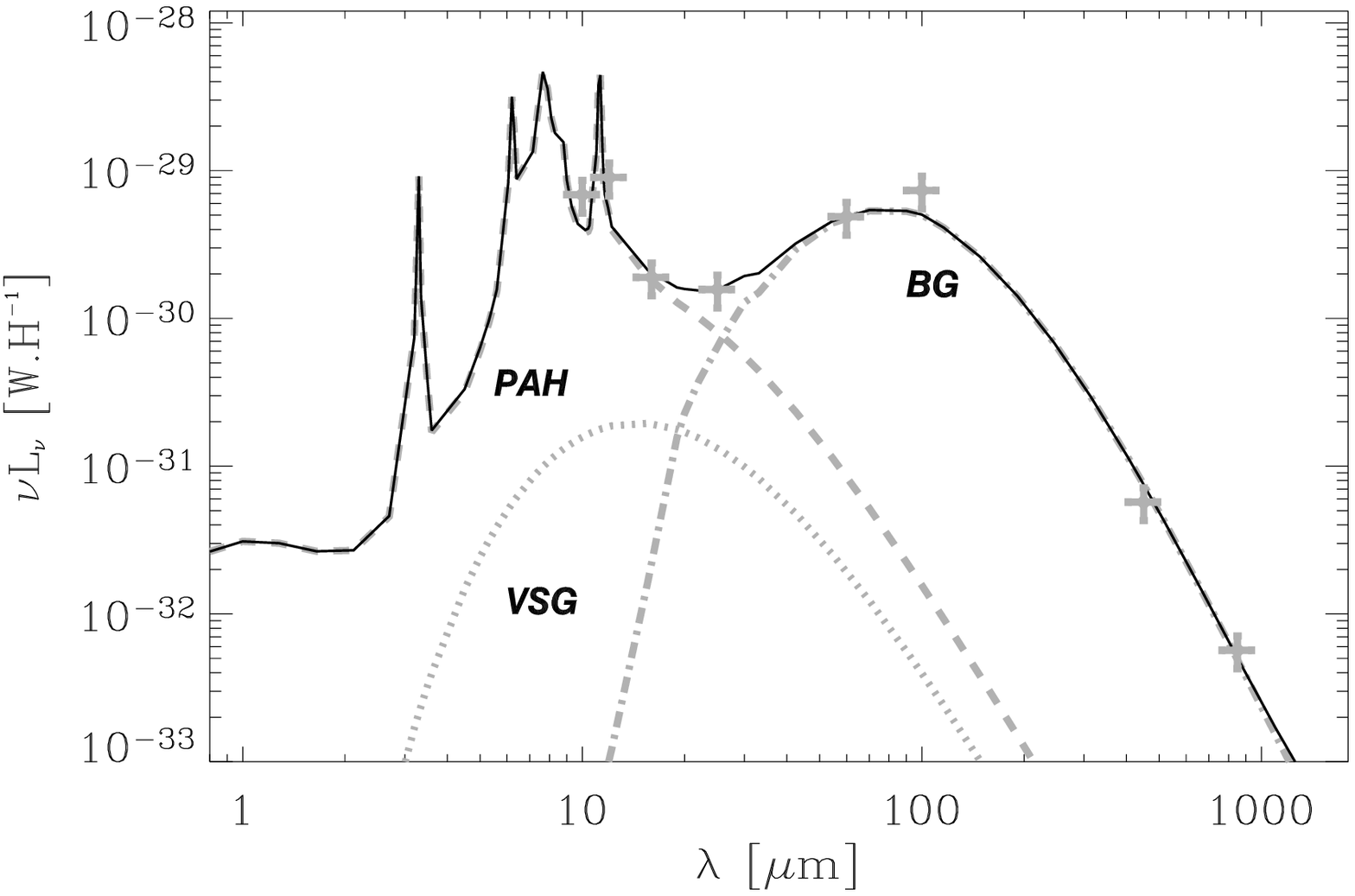}  \\
    \includegraphics[width=0.5\linewidth]{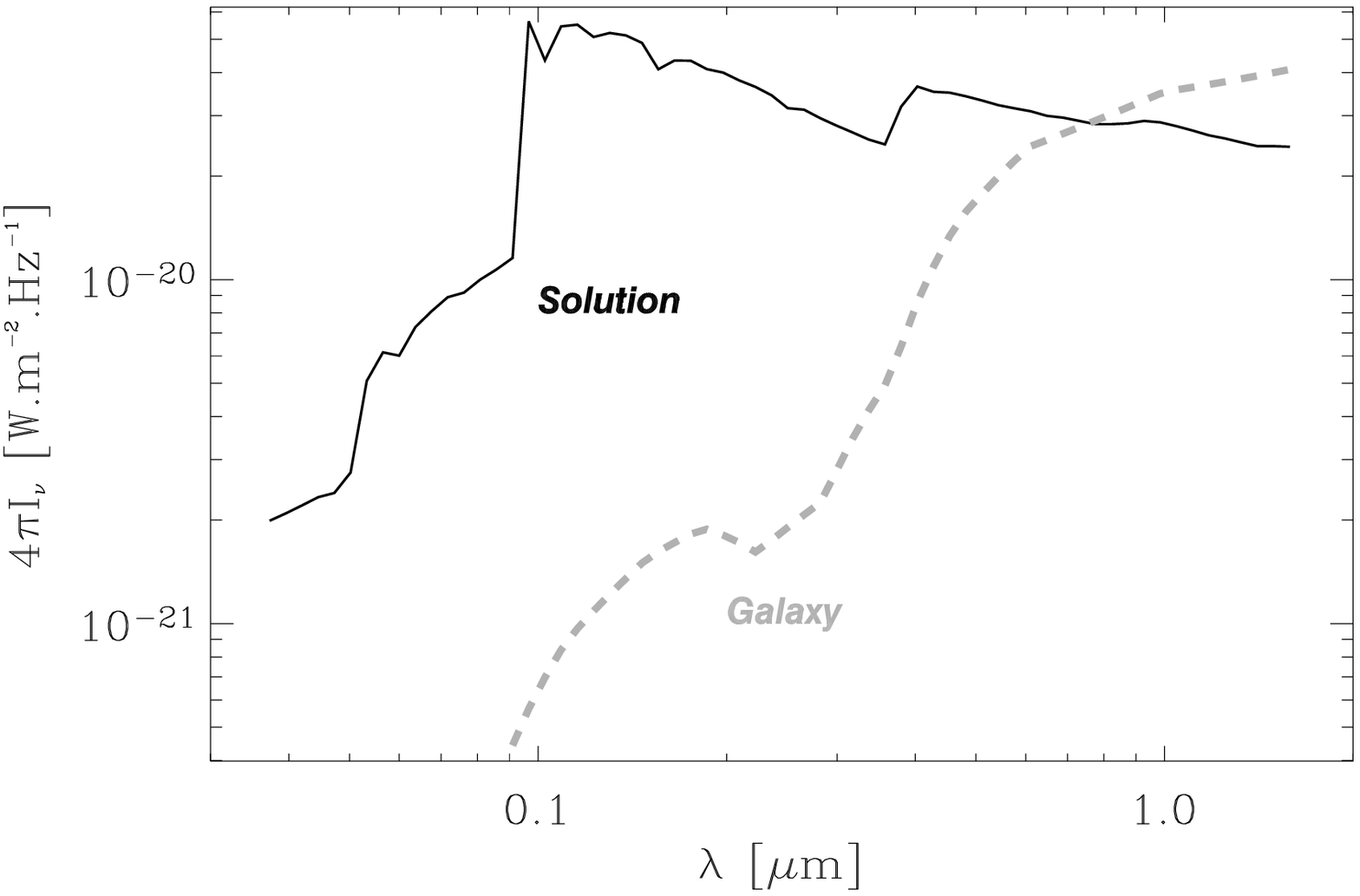} & \hspace*{-0.6cm}
    \includegraphics[width=0.5\linewidth]{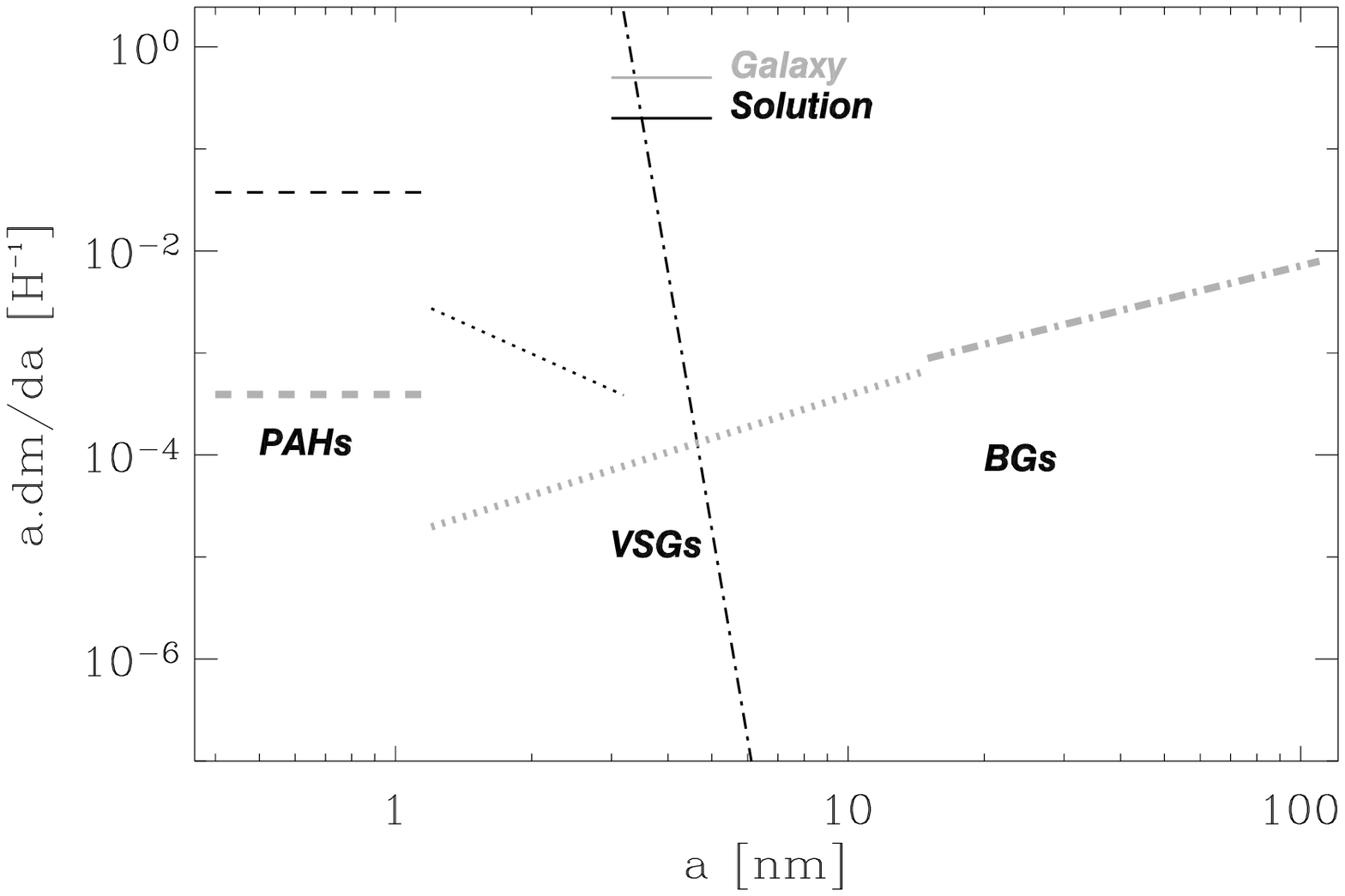} \\
  \end{tabular}
  \caption{Demonstration of the degeneracy in dust SED modelling.
           The model used to fit these SEDs is the \citet{desert+90} model.
           The three dust components are the PAHs (polycyclic aromatic 
           hydrocarbons), the VSGs (very small grains) and the BGs (big 
           grains).
           The grain radius is $a$, and $dm/da$ is the mass of grains
           between $a$ and $a+da$.
           From \citet{galliano+04}.}
  \label{fig:degeneracy}
\end{figure}
Figure~\ref{fig:degeneracy} illustrates the degeneracy between the effects
of the ISRF and those of the size distribution.
The two top plots show an observed dust SED (the grey crosses;
these fluxes are identical in the two plots).
They have been fitted with the \citet{desert+90} model (hereafter DBP90), by 
varying two different sets of 
parameters:\textlist{\thetextlist~the top-left plot has been fitted by 
             varying the ISRF and keeping the Galactic size distribution
           \thetextlist~the top-right plot has been fitted by varying the 
             size distribution and keeping the Galactic ISRF.}
The bottom-left plot shows the radiation field required to produce the
top-left plot, compared to the Galactic one.
This ISRF is harder and more intense, thus the grains are hotter, emitting at
shorter wavelengths.
The bottom-right plot shows the size distribution required for the
top-right plot, compared to the Galactic one.
To fit this SED with a soft and low ISRF, we need to increase the abundance 
of small grains which are the hottest ones.

These figures emphasize the necessity to constrain independently both the 
ISRF and the grain size distribution when modelling a dust SED.

\section{The choice of a simple self-consistent model}

  \subsection{The multiwavelength observations}

\begin{figure}[htbp]
  \centering
  \begin{tabular}{cc}
    \includegraphics[width=0.5\linewidth]{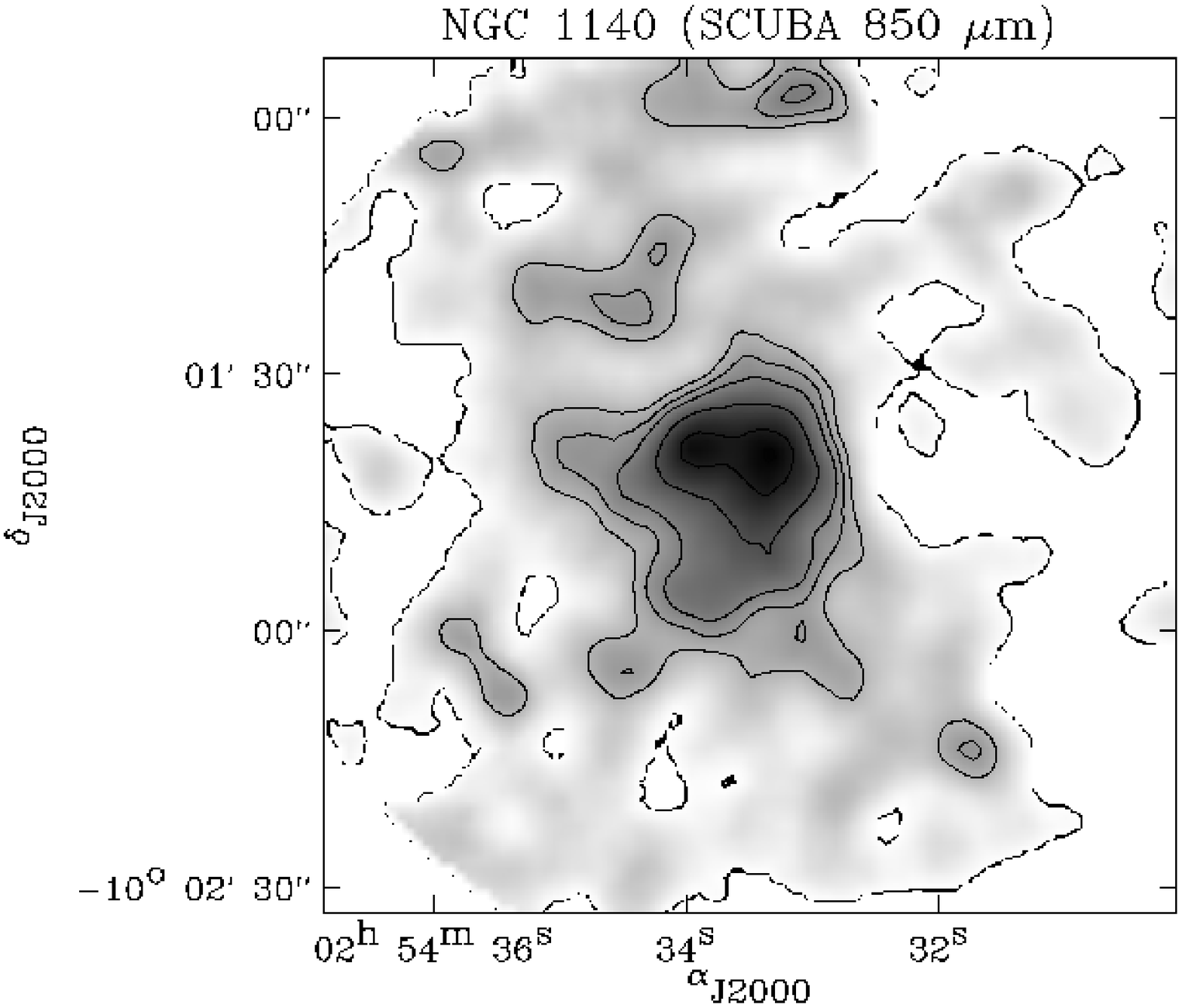}  & \hspace*{-0.6cm}
    \includegraphics[width=0.5\linewidth]{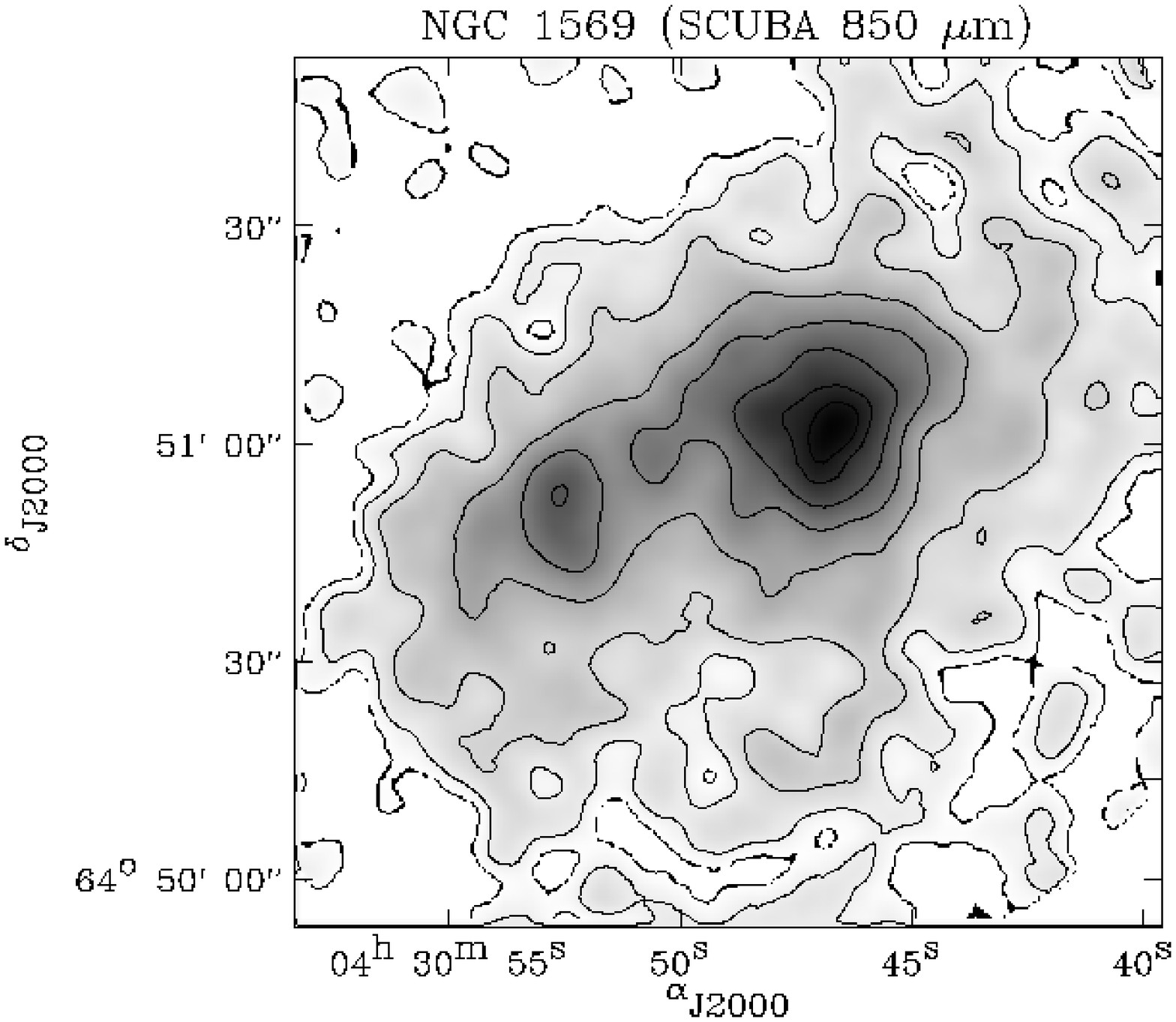}  \\
    \includegraphics[width=0.5\linewidth]{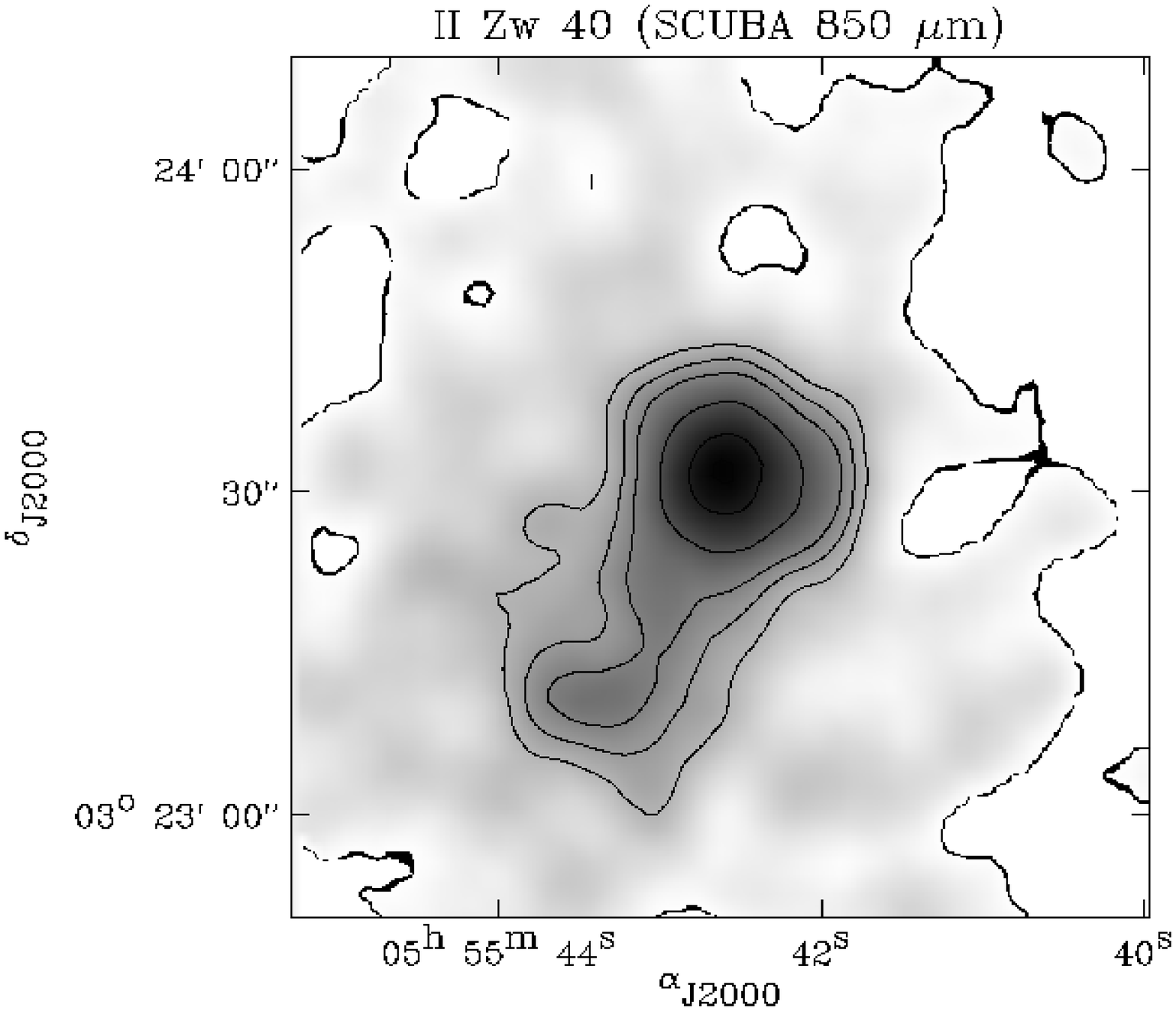}   & \hspace*{-0.6cm}
    \includegraphics[width=0.5\linewidth]{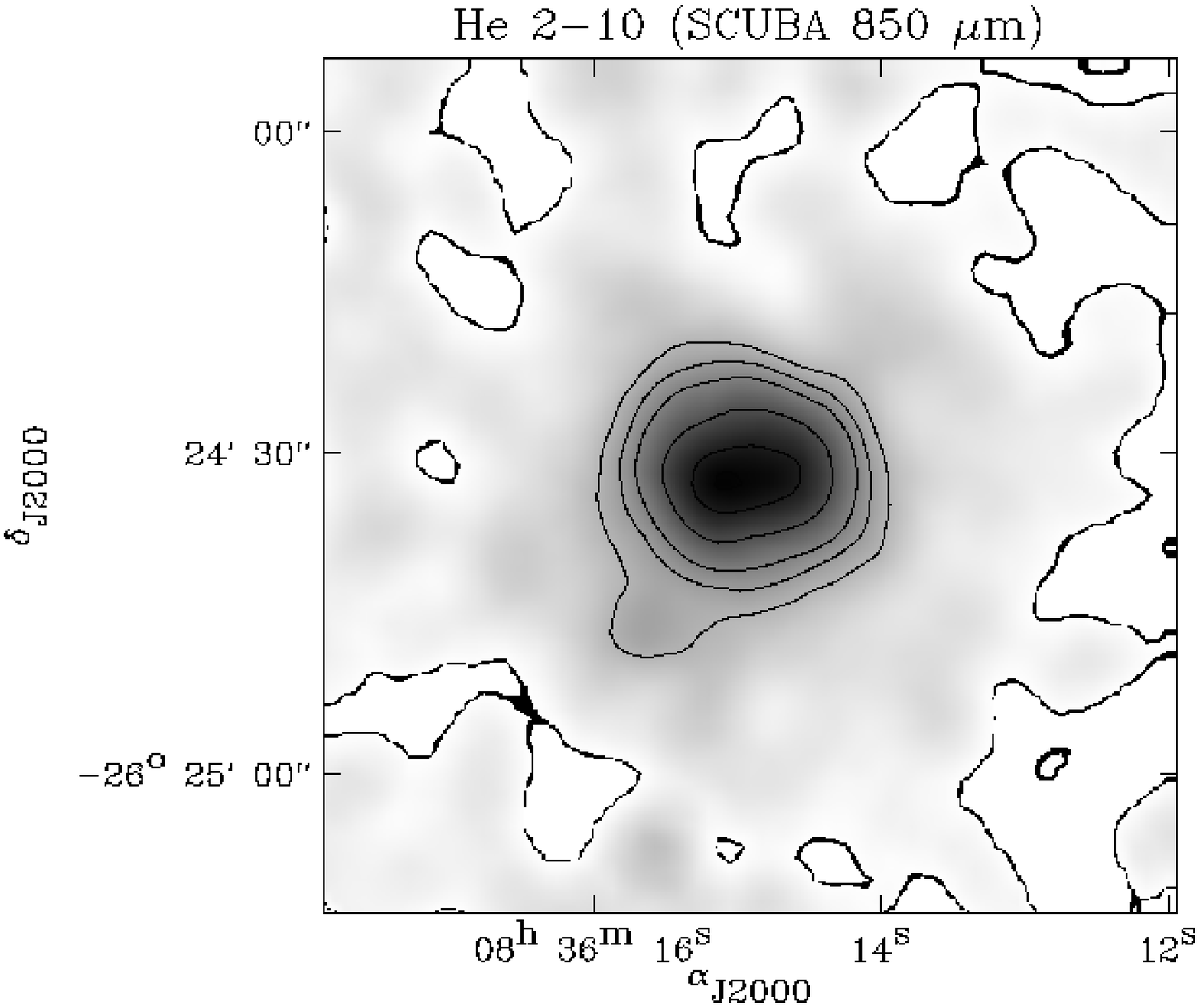}   \\
  \end{tabular}
  \caption{850$\mic$ SCUBA images of the four dwarf galaxies.
           From \citet{galliano+03} and \citet{galliano+04}.}
  \label{fig:scuba}
\end{figure}
We have built the observed UV-to-radio SED of four nearby starburst dwarf 
galaxies (\ngc{1140}, \ngc{1569}, \iizw\ and \hen) by collecting data from the 
literature together with our mid-IR ISOCAM and submm SCUBA 
(Figure~\ref{fig:scuba}) and MAMBO observations.
The broadband fluxes constrain the stellar continuum, the dust emission and
the free-free/synchrotron contribution.
The mid-IR ionic lines observed by ISOCAM are used to constrain the ionized
gas properties, and the CO line measurements are used to remove the molecular 
gas contribution to the submillimeter broadbands.

  \subsection{The model}

The strategy we adopted to model the dust SED of these four nearby dwarf 
galaxies is motivated by the degeneracy shown in Figure~\ref{fig:degeneracy}.
The different steps are the following.
\begin{enumerate}
  \item~An ISRF is synthesized using \peg\ \citep{pegase}.
    We consider two age populations: one accounting for the old 
    underlying stars and a young one accounting for the stars formed 
    during the recent starburst.
    This ISRF fits the unreddened UV-optical broadband fluxes.
    The mean internal optical depth is computed from the energy balance
    between the IR-mm emission and the UV-optical absorption.
    The young stellar population is not well constrained since we lack
    UV data (at shorter wavelengths than U band).
  \item~The ratio of the mid-IR ionic lines (\neiiiline, \neiiline, \sivline) 
    observed by ISOCAM are very sensitive to the hardness of the radiation 
    field and are not significantly affected by extinction.
    The uncertainty of the UV part of the ISRF is removed by constraining
    the photoionisation model \clo\ \citep{cloudy} by these mid-IR ionic line
    ratios.
  \item~The synthesized ISRF is used as the heating source of the 
    DBP90 dust model.
    We vary the grain size distribution in order to fit the IR-mm emission.
    The dust properties deduced from this fit allow us to synthesize an
    average extinction curve for each galaxy.
  \item~This extinction curve is used to unredden the UV-optical observations.
    We iterate these four steps until we reach an agreement between
    extinction and emission properties.
\end{enumerate}

\begin{figure}[htbp]
  \centering
  \includegraphics[width=0.6\linewidth]{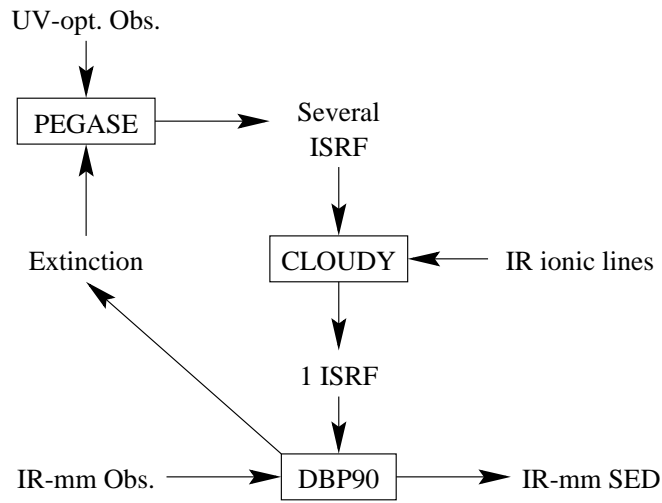}
  \caption{General method used to fit the SEDs of the four dwarf galaxies
           \citep{galliano+03,galliano+04}.}
  \label{fig:method}
\end{figure}
Figure~\ref{fig:method} summarizes the different steps of our modelling.
This method allow us to independently constrain the local ISRF and the average
size distribution in a self-consistent way, taking into account the emission
and extinction properties of the dust.

\section{Average dust properties in four starbursting dwarf galaxies}

\begin{figure}[htbp]
  \centering
  \begin{tabular}{cc}
    \includegraphics[width=0.5\linewidth]{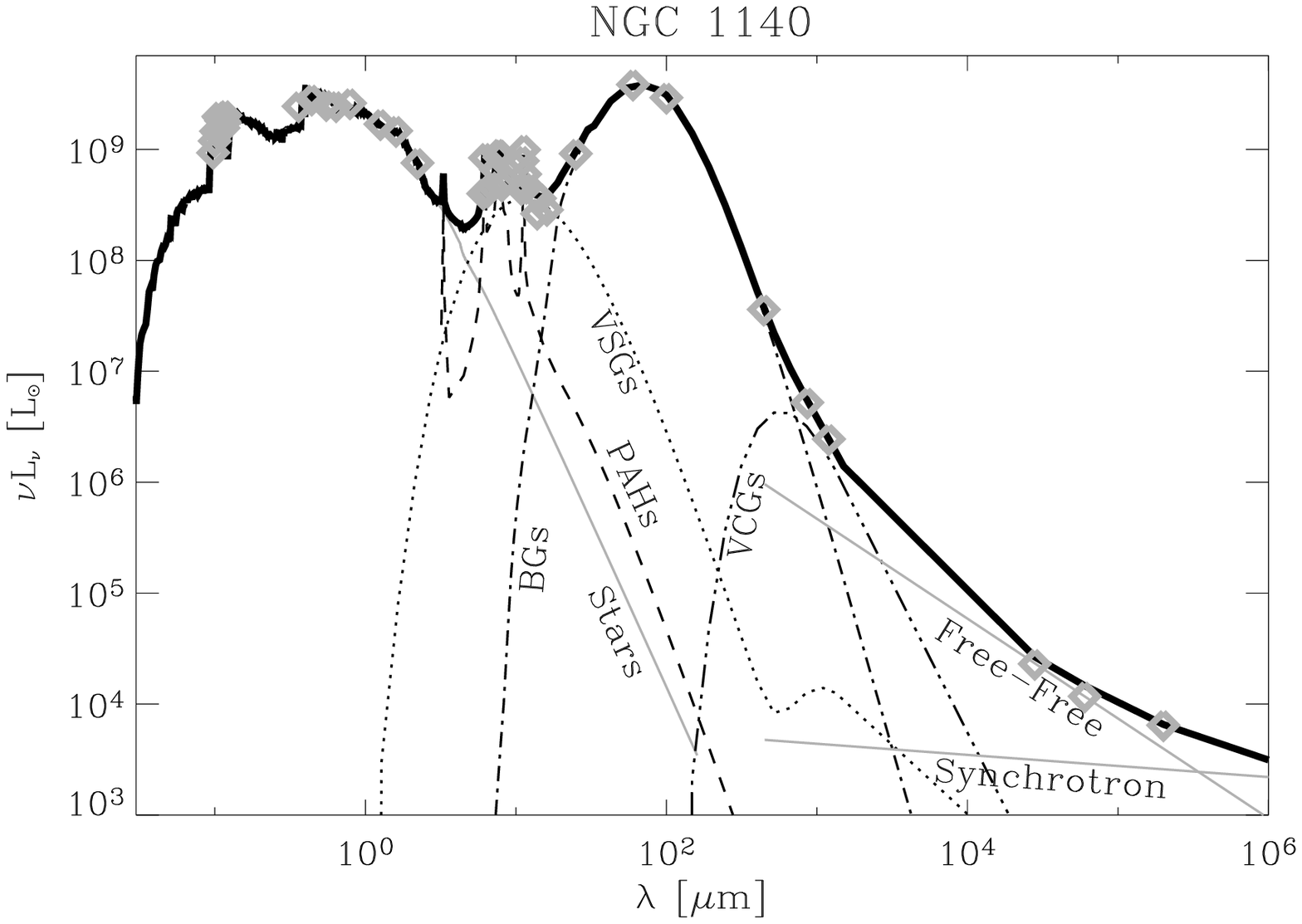} &
    \includegraphics[width=0.5\linewidth]{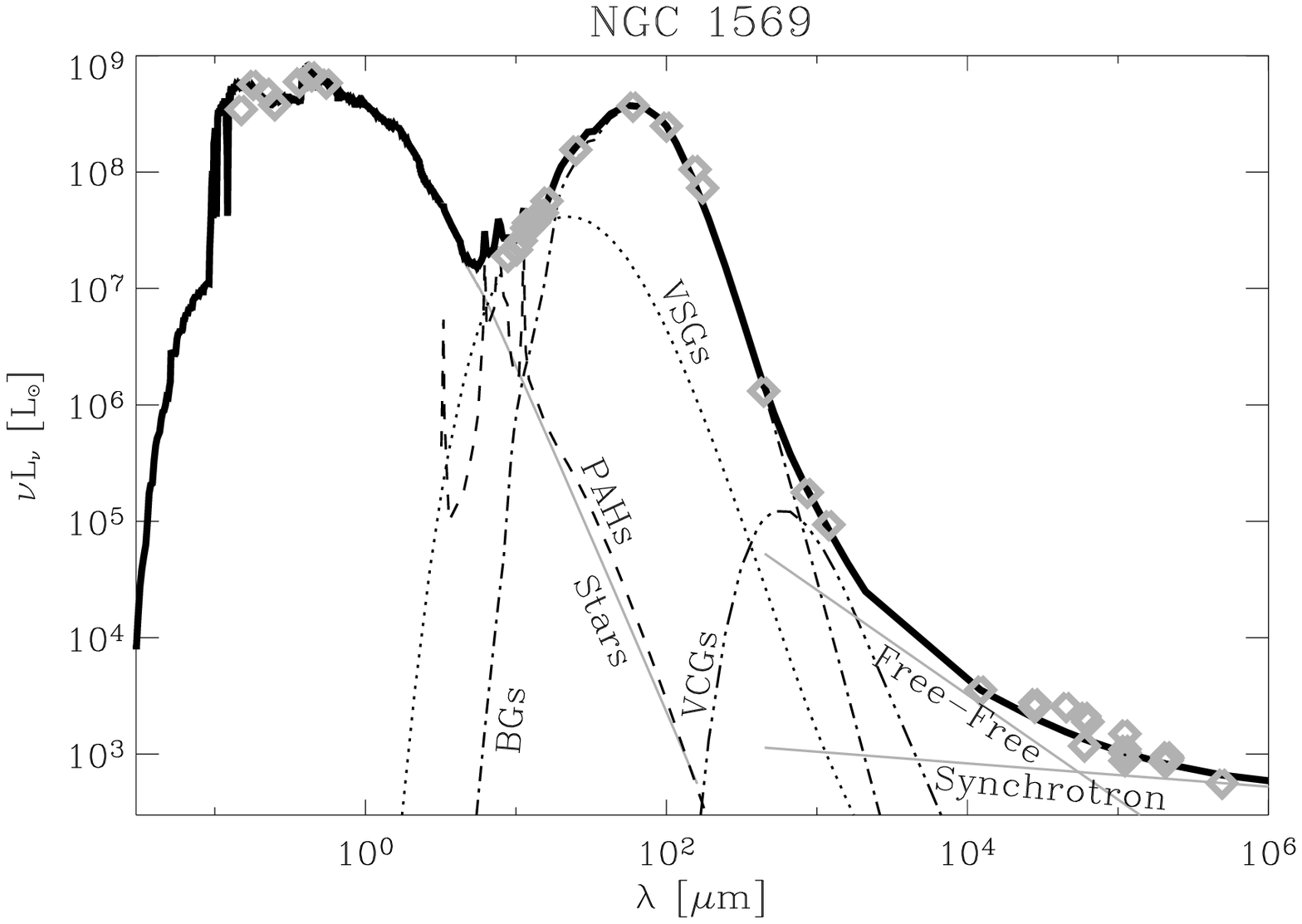} \\
    \includegraphics[width=0.5\linewidth]{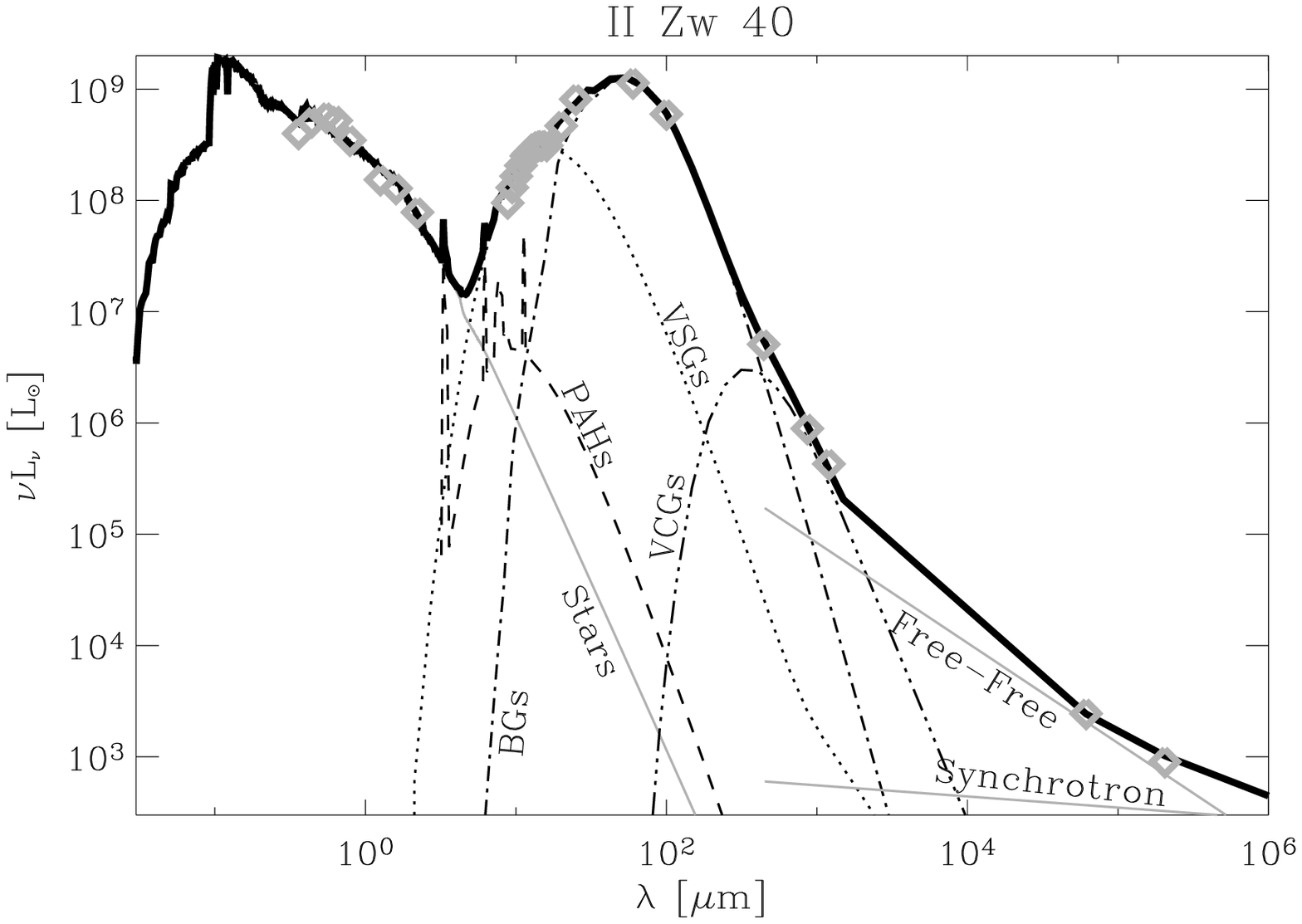} &
    \includegraphics[width=0.5\linewidth]{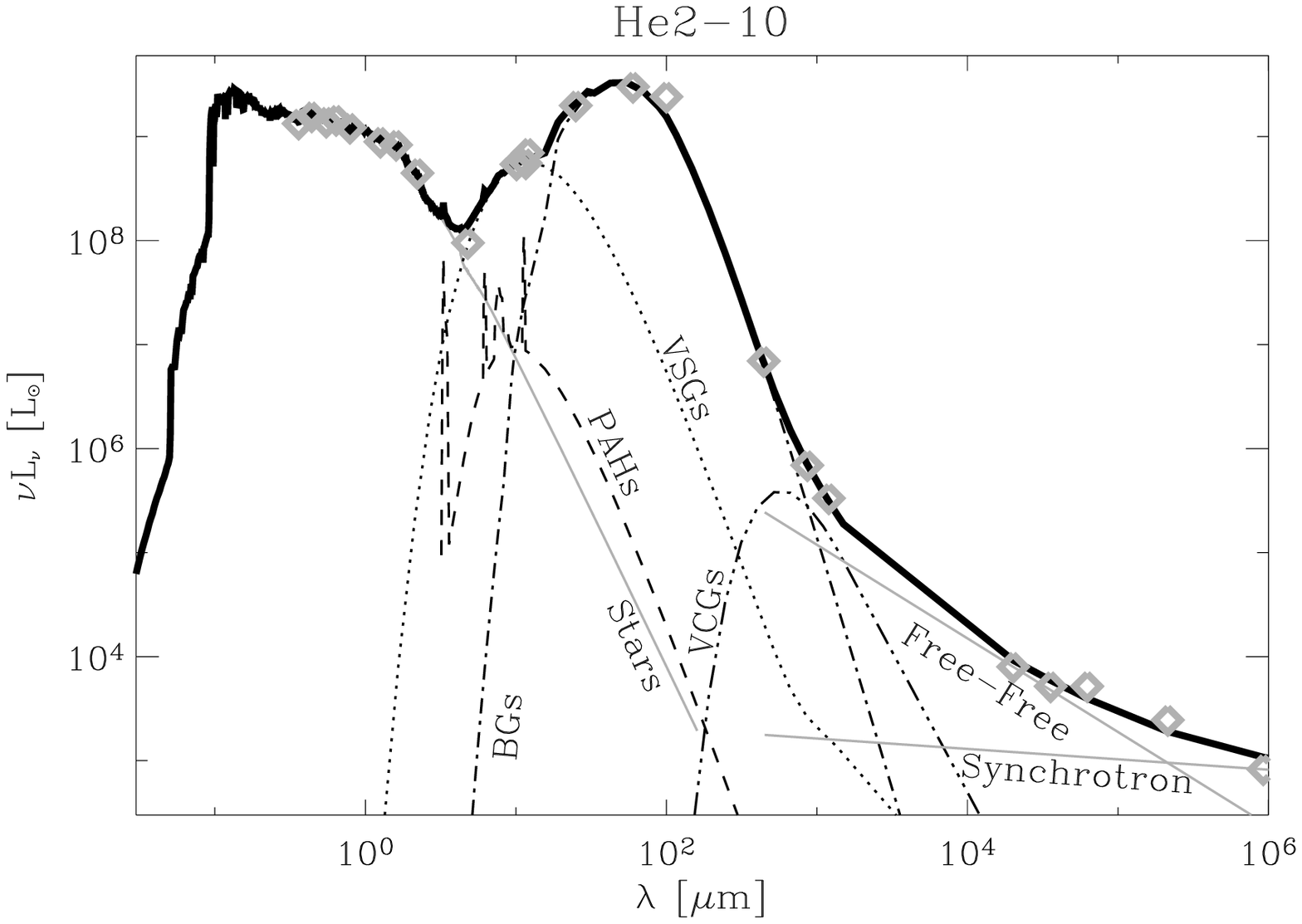}  \\
  \end{tabular}
  \caption{Multiwavelength observed and modelled global SEDs of four dwarf 
           galaxies \citep{galliano+03,galliano+04}.
           The grey diamonds are the observations and the lines are the model.}
  \label{fig:sedtot}
\end{figure}
Figure~\ref{fig:sedtot} shows the global SEDs of our four galaxies.
The IR-mm part, emitted by the dust is hotter than in the Milky Way.
There remain a submillimeter emission excess in each of the SEDs that
we cannot explain with regular dust properties or radio-molecular 
contributions.
In the next section, I will present the modeled dust properties.
Then, I will discuss the consequences of this submillimeter excess.

  \subsection{The modeled dust properties}

We first notice that the PAHs are underabundant in these galaxies.
They are very weak in \ngc{1569}, \iizw\ and \hen.
Only \ngc{1140}, which is the most quiescent of these objects, 
present some relatively strong mid-IR band emission \citep{madden+04}.

\begin{figure}[htbp]
  \centering
  \includegraphics[width=0.6\textwidth]{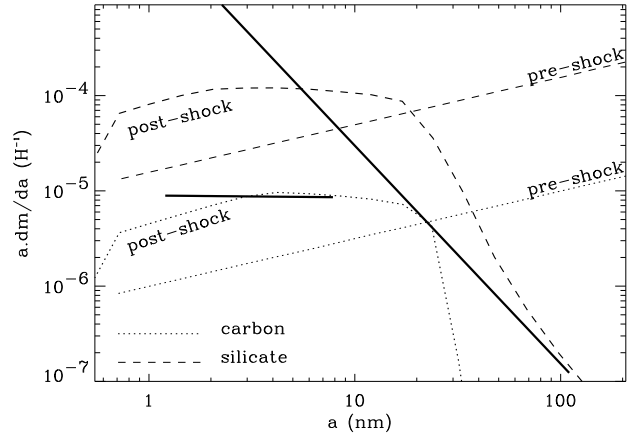}
  \caption{Grain size distribution for \ngc{1569} compared to a graphite 
           and silicate size distribution before (distribution of
           \citep{mathis+77}) and after a shock wave \citep{jones+96}.
           From \citet{galliano+03}.}
  \label{fig:shocks}
\end{figure}

Second, the silicate and graphite size distributions constrained by the fit 
of the dust SED are different from those of the Galaxy.
The grains are on average smaller ($\sim 3-4\,$nm) than those in the 
Milky Way.
Figure~\ref{fig:shocks} shows the size distribution of \ngc{1569} compared to
the Galactic size distribution before and after a shock \citep{jones+96}.
The grain size segregation is consistent with the erosion and the 
fragmentation of the grains by the numerous supernovae shock waves
that these galaxies have recently experienced.

\begin{figure}[htbp]
  \centering
  \includegraphics[width=0.6\textwidth]{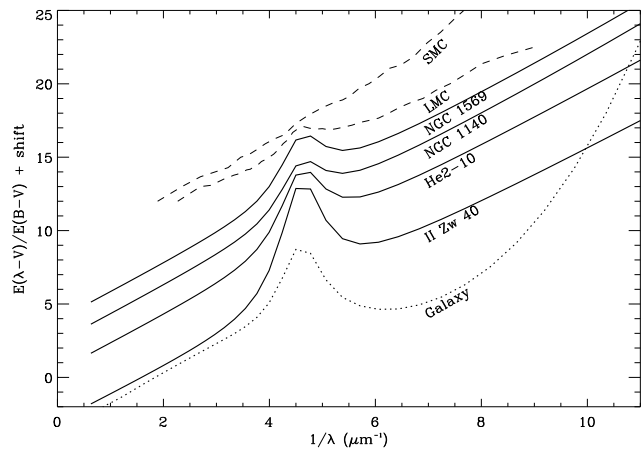}
  \caption{Synthesized extinction curves for our four galaxies compared to the
           Galactic and the Magellanic extinction curves.
           These curves have been shifted for clarity.
           From \citet{galliano+04}.}
  \label{fig:compext}
\end{figure}

A consequence of this small average grain size is that the bulk of the 
dust emission originates into stochastically heated particles.
Another consequence is that the synthesized extinction curves are 
qualitatively similar to those observed toward several lines of sight 
in the Magellanic Clouds.
Their slope is almost linear in $1/\lambda$ and they have a lower 
extinction bump, except in one galaxy (figure~\ref{fig:compext}).

  \subsection{The submillimeter excess}

The submillimeter excess could be either due to peculiar grain optical 
properties or to a component of very cold dust.
Laboratory data show that the emissivity index drops at low temperature
\citep{agladze+96,mennella+98}.
However, these data can not account for this excess, since
the actual change in the slope occurs at a longer wavelength 
\citep{galliano+03}.

\begin{figure}[htbp]
  \centering
  \includegraphics[width=0.6\textwidth]{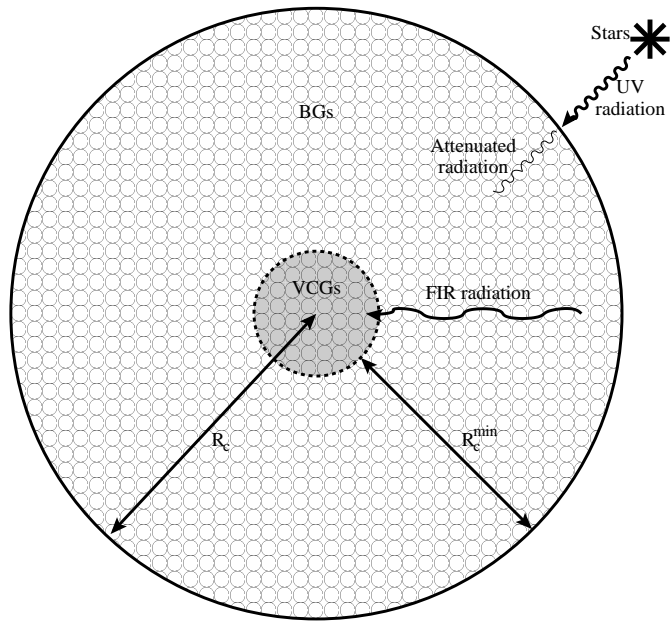}
  \caption{Structure of a clump consistent with the very cold dust hypothesis.
           From \citet{galliano+03}.}
  \label{fig:clump}
\end{figure}

Thus, we explore the second hypothesis which is fully quantifiable.
The temperatures required to fit the excess are low ($T\simeq 5-9\,$K).
The dust mass deduced from this component is a large fraction of the total
dust mass ($40-80\,\%$).
A detailed treatment \citep{galliano+03,galliano+04} shows that this 
very cold grain (VCG) hypothesis is consistent with very small and dense
clumps.
Figure~\ref{fig:clump} shows the structure of such a clump.
The very cold dust is embedded and heated only by the IR radiation.

\section{Prospectives}

This modelling was a first step to investigate the detailed dust properties in 
low-metallicity environments.
A further improvement would be to take into account the geometry of the
environment, by using the spatial information that we have on these galaxies,
and treating the radiative transfer.

The second extension of this study would be to consider a larger sample in
order to disentangle the effects of the low-metallicity of the medium and
the effects of the radiation field.
The knowledge of galaxy SEDs is very important to prepare the future missions
like Herschel.


\begin{theacknowledgments}
  This presentation was performed while the author held a National Research
  Council Research Associateship Award at NASA GSFC.
\end{theacknowledgments}


\bibliographystyle{aipproc}   

\bibliography{../../BibTeX/article,../../BibTeX/techreport}

\IfFileExists{\jobname.bbl}{}
 {\typeout{}
  \typeout{******************************************}
  \typeout{** Please run "bibtex \jobname" to optain}
  \typeout{** the bibliography and then re-run LaTeX}
  \typeout{** twice to fix the references!}
  \typeout{******************************************}
  \typeout{}
 }

\end{document}